\pdfoutput=1
\documentclass[aps,prd,reprint,superscriptaddress,longbibliography,nofootinbib]{revtex4-2}

\usepackage{graphicx}
\usepackage{overpic}
\usepackage{amsmath}
\usepackage{amssymb}
\usepackage{amsfonts}
\usepackage{dcolumn}
\usepackage{dsfont}
\usepackage{latexsym}
\usepackage{rotating}
\usepackage{color}
\usepackage{latexsym}
\usepackage{bbm}
\usepackage{subfigure}
\usepackage{float}
\usepackage{epsfig}
\usepackage{psfrag}
\usepackage{natbib, hyperref}
\usepackage{verbatim}
\usepackage{dcolumn}
\usepackage{makecell}
\newcolumntype{d}[1]{D{.}{.}{#1}}

\usepackage{natbib}
\usepackage{bm}
\usepackage{amsthm}
\usepackage{eucal}
\usepackage{mathrsfs}
\usepackage{url}
\usepackage{braket}
\usepackage{ulem}
\usepackage{calligra}
\usepackage[T1]{fontenc}
\usepackage[utf8]{inputenc}
\usepackage{newunicodechar}
\usepackage{marvosym}
\usepackage[absolute]{textpos}
\usepackage[cal=cm]{mathalfa}
\usepackage{soul}
\usepackage{appendix}

\usepackage{color}

\usepackage{hyperref}
\hypersetup{
colorlinks=true,final=true,
        linkcolor=blue,
        citecolor=blue,
        filecolor=blue,
        urlcolor=blue,
}

\begin{document}
\setlength{\abovedisplayskip}{3pt}
\setlength{\belowdisplayskip}{3pt}

\title{Electronic structure, magnetic order and anisotropic magnetoresistance in $\mathbf{Eu_{1-x}Gd_xCd_2P_2}$ (0 $\le$ x $\le$1) }
\author{Yogendra Singh Negi}
\affiliation{Department of Physics, Indian Institute of Technology Roorkee, Roorkee 247667, India}

\author{Tulika Maitra}
\email[]{tulika.maitra@ph.iitr.ac.in}
\affiliation{Department of Physics, Indian Institute of Technology Roorkee, Roorkee 247667, India}

\begin{abstract}
Recently, $\text{EuCd}_2\text{P}_2$ has attracted increased attention for its colossal magnetoresistive (CMR) property, the mechanism of which seems to deviate from that of known canonical CMR compounds such as manganites. Many studies underscore the role of low density of hole carriers caused by Eu vacancies, particularly in stabilizing ferromagnetic (FM) order and modifying electrical transport and magnetoresistive responses in this compound. A magnetic polaron based mechanism rather than double exchange based mechanism is being proposed to explain the CMR effect for this compound. In this work, we explore the effect of elctron doping on the magnetic transition and CMR behaviour via Gd substitution at Eu sites with varying concentration (x) in Eu$_{1-x}$Gd$_{x}$Cd$_2$P$_2$. Our Density Functional Theory (DFT) and Monte Carlo-based calculations reveal an oscillating ground state magnetic order as a function of electron carrier density. The ground state is found to be A-AFM metal for x=0.056, 0.125 and 0.5 while it becomes ferromagnetic metal at x= 0.33, 0.67 and 1. Most importantly, we predict colossal anisotropic magnetoresistance in this compound for x=0.125 and 0.5. The underlying mechanism of anisotropic CMR in this compound looks similar to double exchange like due to the presence of a strong intra-site 4\textit{f}-5\textit{d} ferromagnetic exchange interaction and an itinerant electron in the hybridized conduction states. The oscillating ground state magnetic order and the enhancement of transition temperature (as high as 150K for x=0.67) as a function of electron carrier density offers a new avenue to tune the magnetotransport property in this compound and offers an excellent platform for designing future spintronics devices.                       
\end{abstract}

\maketitle
\section{\textbf{INTRODUCTION}}
\vspace{-2mm}
Eu based 122 compounds (Eu$T_2X_2$, $T$ = Cd, Zn; $X$ = P, As, Sb) with $\text{CaAl}_2\text{Si}_2$ type structure have drawn huge interest, for hosting diverse phenomena like magnetic exchange driven topological properties, Colossal Magnetoresistance (CMR), giant thermopower and so on~\cite{Mp_EuCd2P2_II,Weyl_EuCd2As2,EuCd2Sb2,Thermo_EuZn2Sb2,EuZn2As2,EuCd2As2_Ist,CMR_EuZn2P2,CMR_EuCd2As2,EuCd2As2_MR,paper_EuCd2P2,Giant_Th_Power_EuCd2P2}. Among these compounds, EuCd$_2$P$_2$ and EuZn$_2$P$_2$ exhibit CMR behavior~\cite{paper_EuCd2P2,CMR_EuZn2P2}. Contrary to most CMR compounds which exhibit FM ground state, EuCd$_2$P$_2$ and EuZn$_2$P$_2$ have A-type AFM (A-AFM) ground state where ferromagnetic ab-planes are stacked antiferromagnetically along the $c$-direction. In EuCd$_2$P$_2$, the antiferromagnetic transition is observed at $T_\text{N}$ = 11 K and resistivity peak at 18 K~\cite{paper_EuCd2P2,HeatCapacity_EuCd2P2}. Furthermore, recent reports reveal that the formation and percolation of ferromagnetic clusters (or magnetic polarons) drive the underlying CMR behavior~\cite{Sunkoetal_EuCd2P2,Mp_EuCd2P2,Mp_EuZn2P2,Mp_EuCd2P2_II}, mirroring mechanisms observed in classic FM CMR systems such as Europium hexaboride, Europium chalcogenides etc.~\cite{EuB6_Mp,EuB6_Mp_II,EuO_Mp,EuO_Mp_II}, unlike manganites where double exchange mechanism coupled with Jahn-Teller distortion drive the CMR~\cite{Manganite_CMR,Manganite_review_CMR}.\par

The intriguing properties of the Eu$T_2X_2$ family stem from the interplay between its narrow electronic band gap, low carrier density, and local Eu magnetic moment~\cite{EuT2X2_review}. Recently, Chen \textit{et al.}~\cite{hole_doped_EuCd2P2} demonstrated that the magnetic ground state of $\text{EuCd}_2\text{P}_2$ can be tuned from semiconducting A-AFM ($T_\text{N}$ = 11 K) to metallic FM ($T_\text{C}$ = 47 K) state by introducing a very low hole concentration, corresponding to 0.26\% Eu vacancies per formula unit (f.u.) in contrast to the nearly 0.02\% Eu vacancy per f.u. in the AFM state. In fact, similar behavior has also been observed in sister compound EuZn$_2$P$_2$~\cite{hole_doped_EuZn2P2}, where varying hole doping generates a carrier concentration in the range of $10^{19}$ to $10^{21}\ \text{cm}^{-3}$ with varying FM transition temperatures ($T_\text{C}$)~\cite{EuT2X2_review}. This wide range of carrier density stems directly from the diverse crystal growth techniques. Thus, the tunability of magnetic and transport properties by variation of carrier concentration in these compounds offers a new avenue for their potential use in spintronics applications. 
Furthermore, the layered structure of Eu$T_2X_2$ facilitates the fabrication of heterostructures and thin films, making it an excellent candidate for future spintronics technology~\cite{paper_EuCd2P2}.


In Eu-based 122 systems, most of the experimental and theoretical work reported so far in the context of CMR, have looked into the hole doped regime as discussed above. However, the effect of electron doping can be interesting as well and can show very different effects. For example, in manganites, the mechanism of CMR in the hole and electron doped regimes is found to be very different~\cite{Hole_doped_Manganite,TM_2002,TM_2003_PRB,Ce_doped_manganite}. Therefore, it would be very interesting to investigate the effect of electron doping on the CMR mechanism in this compound. It is well known that the magnetic and transport properties can be effectively tuned over a wide range through external pressure, carrier doping, and chemical substitution while stoichiometric changes offer only a limited variation in the carrier concentration~\cite{EuT2X2_review}. Chemical substitution has been seen to provide the desirable wide range of carrier densities in Europium hexaboride~\cite{EuGdB6_RKKY}, Europium chalcogenides~\cite{EuX_reviewpaper,GdEuSe_GMR,Gd_doped_EuO}, and $\text{EuCd}_2\text{As}_2$~\cite{Na_substituted_EuCd2As2}. 

In the present work, we systematically investigate the magnetic and transport properties of electron-doped $\text{EuCd}_2\text{P}_2$ via isovalent $\text{Gd}$ substitution at the Eu site at the varying concentration to elucidate the precise role of carrier density in engineering the magnetic transition temperature. Gd has been chosen as a viable dopant as it has electronic configuration 4${f}^7$5${d}^1$ in 2+ oxidation state compared to the 4${f}^7$5${d}^0$ of Eu$^{2+}$. Consequently, this offers a two-fold advantage, such as no magnetic dilution effect due to the identical half filled $f$-orbitals, and at the same time it injects an extra electron to the lattice via the occupied 5\textit{d}-orbital.

\begin{figure}[h!]
	\centering
	\includegraphics[width=8.5cm]{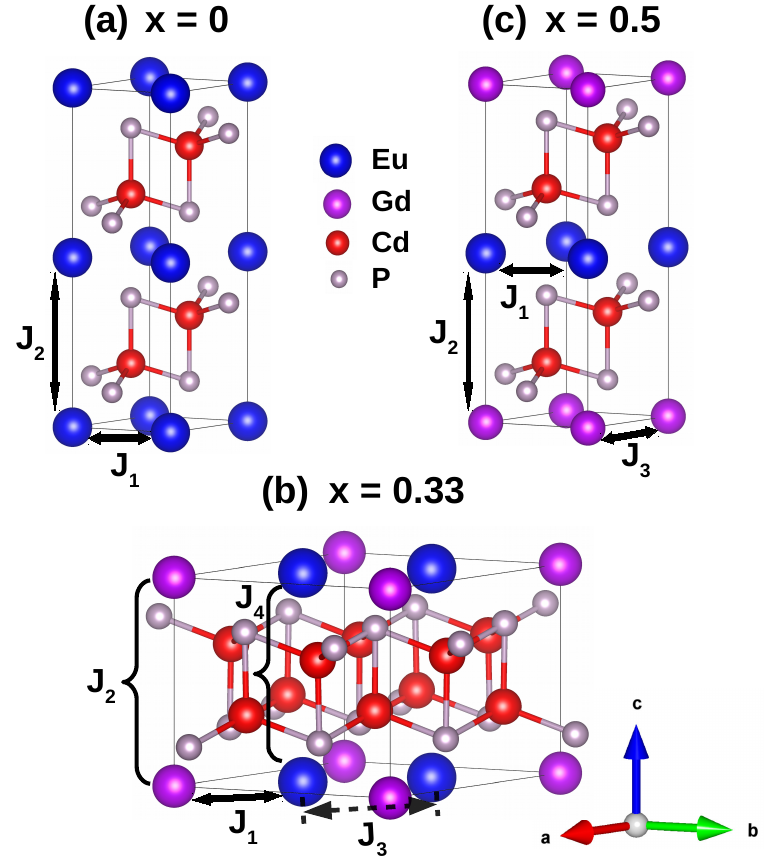}
	\caption{Crystal structure of (a) EuCd$_{2}$P$_{2}$, (b) Eu$_{0.67}$Gd$_{0.33}$Cd$_{2}P_{2}$ and (c) Eu$_{0.5}$Gd$_{0.5}$Cd$_{2}$P$_{2}$, where Gd, Eu, Cd and P ions are shown in magenta, blue, red and grey colors respectively. The exchange interactions considered in the calculations are also shown for each case.}
	\label{Fig1}
\end{figure}


The paper is arranged as follows. In Sec.~\ref{SecII}, we discuss the details of methodology and parameters used in this work. In Sec.~\ref{SecIII}, main results are discussed in detail and Sec.~\ref{SecIV} is devoted to discussion. Finally, we summarize our results and discuss the future possibility for fostering our understanding of Colossal Magnetoresistance in these compounds.

\section{\textbf{METHODOLOGY}}
\label{SecII}
\vspace{-2mm}
To obtain the total energies and electronic structures of Eu$_{1-\text{x}}$Gd$_\text{x}$Cd$_2$P$_2$, density functional theory (DFT) calculations were performed using projector-augmented wave (PAW)~\cite{PAW_method} method, as implemented in Vienna \textit{ab initio} simulation package ({\small{VASP}})~\cite{VASP}. Perdew–Burke–Ernzerhof generalized gradient approximation (PBE-GGA)~\cite{PBE} was employed for the exchange correlation functional. Coulomb correlation ($U$) and spin-orbit (SO) interaction were also included within GGA+$U$ and GGA+$U$+SO approximations. In all GGA+$U$ and GGA+$U$+SO calculations, $U_{\text{eff}}$ = $U$-$J$ = 7.0 eV~\cite{DFT_EuCd2As2_correct} (where $U$ is Coulomb correlation and $J$ is Hund’s exchange) was included to account for the onsite Coulomb interaction of Eu-4$f$ and Gd-4$f$ orbitals within the rotationally invariant method by Dudarev \textit{et al.}~\cite{GGA+U} and cutoff energy of the plane wave basis was set to 400 eV. For the parent compound (x = 0), $21\times21\times13 $ $\Gamma$-centered \textit{k} mesh was used to perform integration over the Brillouin zone. The experimental lattice parameters~\cite{paper_EuCd2P2} were used, and only the atomic positions were optimized with force and energy convergence criteria of $10^{-2}$ eV/\AA~ and $10^{-6}$ eV, respectively. Subsequently, DFT calculations were performed for several Gd doping concentrations (i.e., 0.056, 0.125, 0.33, 0.5, 0.67, and 1). In these cases, both the lattice constants and atomic positions were fully optimized with an appropriate $\Gamma$-centered \textit{k}-point mesh corresponding to the chosen supercell (i.e., $4\times4\times2$, $11\times11\times7$, $11\times11\times13$, $21\times21\times7$, $11\times11\times13$, and $21\times21\times13$, respectively) using the same aforementioned convergence criteria. To determine the dynamical stability of Eu$_{1-\text{x}}$Gd$_\text{x}$Cd$_2$P$_2$ for different Gd concentrations, phonon dispersion calculations were performed using density functional perturbation theory (DFPT), as implemented in {\small{VASP}} and the {\small{PHONOPY}} package~\cite{DFPT_1,DFPT_2}. For each composition, an appropriate supercell was adopted based on the corresponding lattice parameters. Monte-Carlo simulations (as implemented in the Uppsala Atomistic Spin Dynamics ({\small{UppASD}}) software~\cite{UppAsd_MC}) were performed on the Heisenberg spin model to calculate the magnetic transition temperatures of Eu$_{1-\text{x}}$Gd$_\text{x}$Cd$_2$P$_2$ at various Gd concentration values. 

\section{\textbf{Calculations and Results}}
\label{SecIII}
\vspace{-2mm}
As discussed in the introduction, the effect of hole doping on EuCd$_2$P$_2$ via creation of Eu vacancies has been studied mostly experimentally. In order to investigate the electron doping effect systematically, we substituted Eu with Gd in the supercell of EuCd$_2$P$_2$ to achieve different electron carrier concentrations (see Fig.~\ref{Fig1}). We have studied the entire series  Eu$_{1-\text{x}}$Gd$_\text{x}$Cd$_2$P$_2$ for x = 0, 0.056, 0.125, 0.33, 0.5, 0.67 and 1. We discuss below in detail the results for x = 0, 0.33, 0.5, 0.67 and 1. The dynamical stability of all these compounds are confirmed by phonon calculations where we do not see any soft phonon modes (see Fig.~\ref{Fig2}) which indicates the feasibility of experimental preparation of these compounds. 

\subsection{$\mathbf{EuCd_2P_2}$: \textbf{Parent Compound}}
\vspace{-2mm}


\begin{figure*}[htbp!]
	\centering
	\includegraphics[width=17cm]{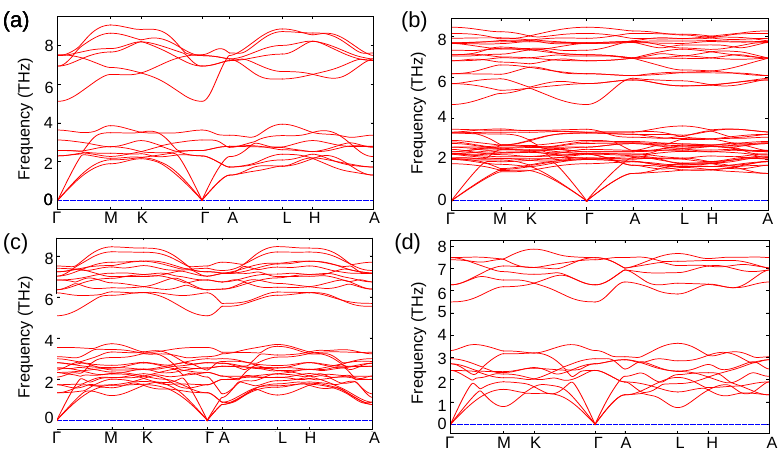}
	\caption{Phonon band structures of Eu$_{1-\text{x}}$Gd$_\text{x}$Cd$_2$P$_2$. Panels (a), (b), (c) and (d) correspond to \text{x} = 0, 0.33, 0.5 and 1, respectively.}
	\label{Fig2}
\end{figure*}


As discussed in the methodology section, we have relaxed the crystal structure of the parent compound $\text{EuCd}_2\text{P}_2$ and performed phonon calculations to ensure its dynamical stability. Our phonon calculations confirms the dynamical stability of $\text{EuCd}_2\text{P}_2$, as we do not see any soft phonon modes in phonon frequencies shown in Fig.~\hyperref[Fig2]{\ref{Fig2}(a)}. We performed the total energy calculations within GGA+\textit{U} and GGA+\textit{U}+SO approximations to establish the ground state magnetic order. As we can see in Table~\ref{TableI} that A-AFM with Eu magnetic moments pointing along $a$ direction (A-AFM [100])) is the magnetic ground state which is consistent with experimental observations~\cite{paper_EuCd2P2, HeatCapacity_EuCd2P2}. \par

\begin{figure*}[t]
	\centering
	\includegraphics[width=17cm]{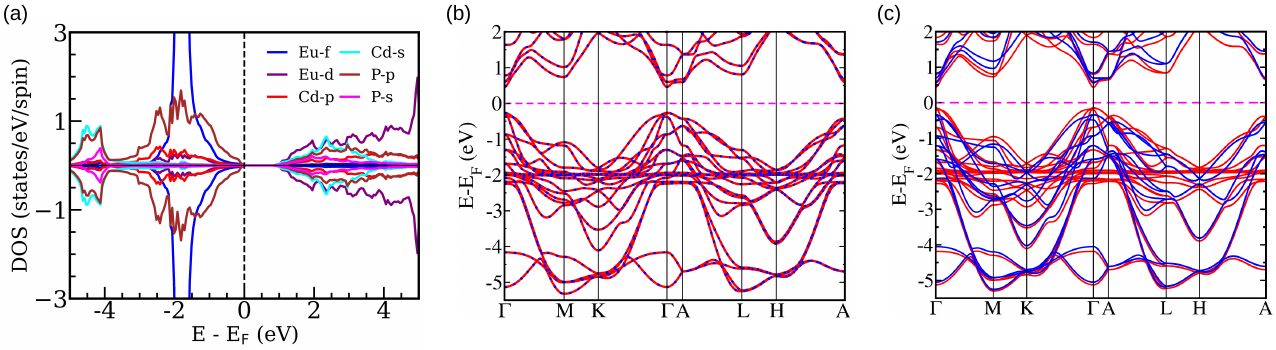}
	\caption{Electronic structure of $\text{EuCd}_2\text{P}_2$ calculated within the $\text{GGA} + U$ approximation: (a) partial density of states (PDOS) in the A-AFM state, (b) band structure in the A-AFM state, and (c) band structure in the FM state. The red and blue lines represent spin-up and spin-down states, respectively.}
	\label{Fig3}
\end{figure*}

From our electronic structure calculations, $\text{EuCd}_2\text{P}_2$ is found to be a semiconductor with a band gap of 724 meV and 583 meV {without SOC} in its A-AFM and FM states, respectively, as one can see in the DOS and band structures presented in  Figs.~\hyperref[Fig3]{\ref{Fig3}(a)},~\hyperref[Fig3]{\ref{Fig3}(b)}, and ~\hyperref[Fig3]{\ref{Fig3}(c)}. In order to simulate the paramagnetic state, we performed calculations with Eu moments frozen as core states and we find a substantially larger band gap of about 840 meV (without SOC) (see Fig. S1 in supplemental material). As the band gap is seen to decrease while going from paramagnetic to A-AFM and then to FM state, this correlates well with the experimental observation of magnetoresistance in this system~\cite{HeatCapacity_EuCd2P2,Band_reconsruct_EuCd2P2,EuCd2P2_GS_related}. Moreover, a similar trend is observed in the band gap upon inclusion of SOC (see Fig. S1 in Supplemental Materials).

We further calculated various exchange constants (\textit{J}) for $\text{EuCd}_2\text{P}_2$ by energy map analysis~\cite{J_Calculation_ref,Superexchange_EuZn2P2}. We calculated all energies within GGA+\textit{U}+SO approximation with spin axis along [100]-direction. Four magnetic configurations have been used to calculate the J values (i.e., A-AFM, FM, A-AFM with one spin reversed (AFM$^\prime$) and FM with one spin reversed (FM$^\prime$). Here, $J_1$ and $J_2$ denote the exchange constants for the intralayer Eu-Eu (nearest neighbor) and interlayer Eu-Eu interactions as shown in Fig.~\hyperref[Fig1]{\ref{Fig1}(a)}. We assumed S = 1 for all spins in a magnetic configuration for $J$ calculation. The calculated $J$ values are listed in Table~\ref{TableII}. 

\begin{figure*}[t]
	\centering
	\includegraphics[width=17cm]{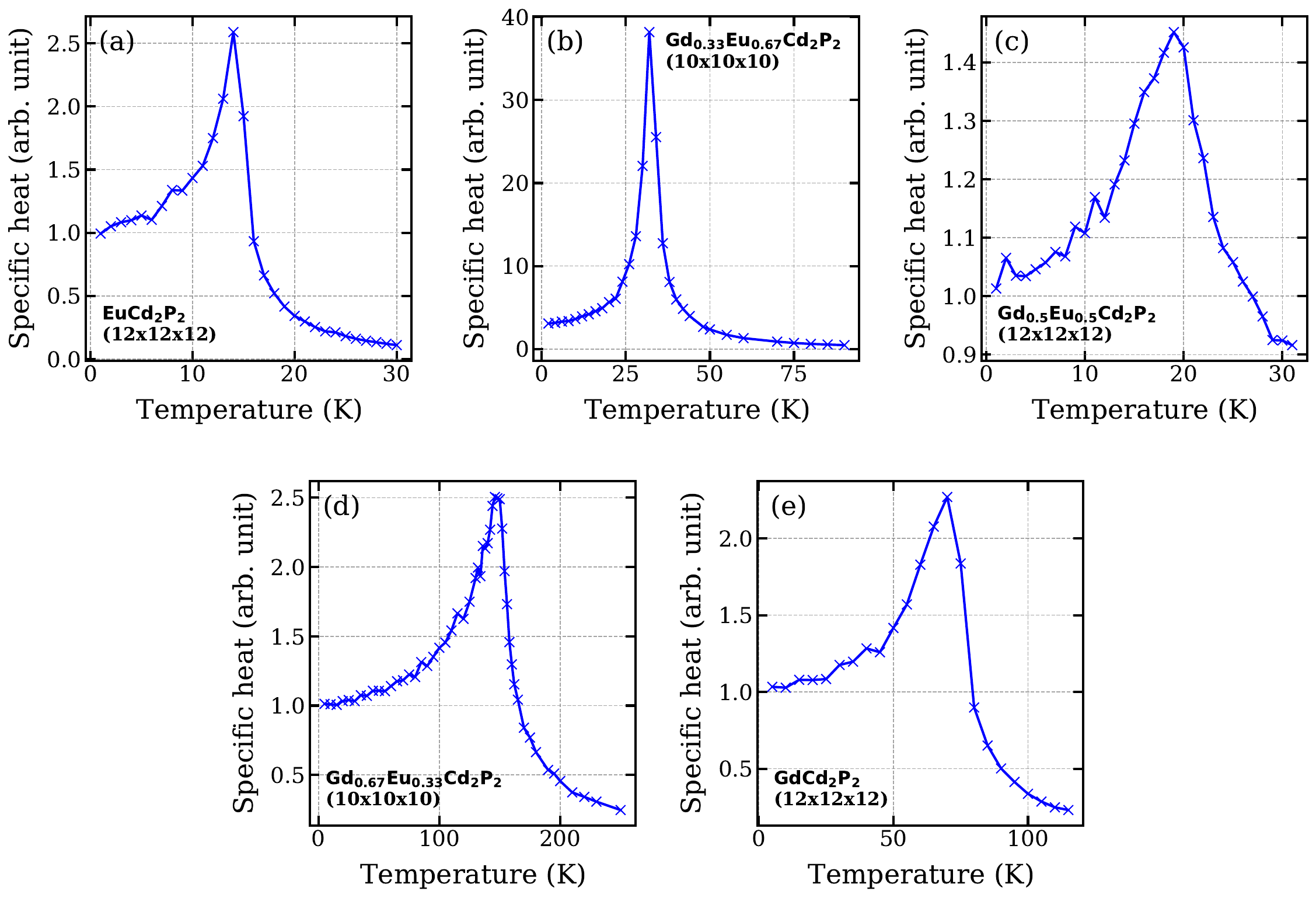}
    \caption{Transition temperature of Eu$_{1-\text{x}}$Gd$_{\text{x}}$Cd$_{2}$P$_{2}$ for: (a) EuCd$_{2}$P$_{2}$, (b) Gd$_{0.33}$Eu$_{0.67}$Cd$_{2}$P$_{2}$, (c) Gd$_{0.50}$Eu$_{0.50}$Cd$_{2}$P$_{2}$, (d) Gd$_{0.67}$Eu$_{0.33}$Cd$_{2}$P$_{2}$, and (e) GdCd$_{2}$P$_{2}$.}
	\label{Fig4}
\end{figure*}

Considering these $J$ values, we performed Monte-Carlo simulations on a Heisenberg model to find the transition temperature. The A-AFM transition temperature ($T_\text{N}$) of $\text{EuCd}_2\text{P}_2$ was found to be 14 K, as shown in Fig.~\hyperref[Fig4]{\ref{Fig4}(a)}, where we have presented the specific heat ($C_\text{V})$) versus Temperature (T) plot. This transition temperature is very close to the experimentally obtained transition temperature of 11 K~\cite{paper_EuCd2P2, HeatCapacity_EuCd2P2,Mp_EuCd2P2}.

\begin{table*}[t]
	\centering
	\caption{\label{TableI}Calculated energy differences [$\Delta E = E_{\text{A-AFM}} - E_{\text{FM}}$, in meV/f.u.] for various Gd concentrations ($\text{x}$) under different approximations. A positive $\Delta E$ indicates that the ferromagnetic state is energetically favored and M is magnetization direction. }
	\vspace{4pt} 
	\renewcommand{\arraystretch}{1.2} 
	\begin{ruledtabular}
		\begin{tabular}{l *{3}{d{2.4}} c}
			\noalign{\vskip 4.0pt} 
			\makecell[l]{$\text{Eu}_{\text{1-x}}\text{Gd}_{\text{x}}\text{Cd}_{2}\text{P}_{2}$} & 
			\multicolumn{1}{c}{\makecell{GGA+$U$ \\ $[\Delta E]$}} & 
			\multicolumn{1}{c}{\makecell{GGA+$U$+SO \\ $[\text{M} \parallel \text{a}$, $\Delta E]$}} & 
			\multicolumn{1}{c}{\makecell{GGA+$U$+SO \\ $[\text{M} \parallel \text{c}$, $\Delta E]$}} & 
            \multicolumn{1}{c}{\makecell{MAE\footnotemark[1] \\ {[meV/f.u.]}}} \\
 			\noalign{\vskip 4.0pt} 
			\colrule
			$\text{x} = 0$   & -0.713 & -0.704 & -0.702  & 0.025 \\
			$\text{x} = 0.056$ & -0.374 & \multicolumn{1}{c}{---} & \multicolumn{1}{c}{---} & \multicolumn{1}{c}{---} \\
			$\text{x} = 0.125$ & -0.349 & \multicolumn{1}{c}{---} & \multicolumn{1}{c}{---} & \multicolumn{1}{c}{---} \\
			$\text{x} = 0.33$ &  4.733 &  4.730 &  4.766  & 0.098 \\
			$\text{x} = 0.5$  & -0.110 & -0.060 & -0.059 & 0.043 \\
			$\text{x} = 0.67$ &  6.416 &  6.270 &  6.230  & 0.185 \\
			$\text{x} = 1$   &  4.039 &  4.078 &  4.085  & 0.120 \\
		\end{tabular}
	\end{ruledtabular}
     \footnotetext[1]{Magneto-crystalline anisotropy energy between the [001] and [100] directions for the ground state.}
\end{table*}
\begin{table}[htbp!]
    \caption{\label{TableII}Calculated exchange parameters ($J$, in meV) for representative Gd concentrations. Positive values denote ferromagnetic interactions.}
	\vspace{4pt}
	\renewcommand{\arraystretch}{1.2}
	\begin{ruledtabular}
		\begin{tabular}{l *{4}{d{2.4}}}
			\noalign{\vskip 3pt}
			\text{Compound} & 
			\multicolumn{1}{c}{$J_1$} & 
			\multicolumn{1}{c}{$J_2$} & 
			\multicolumn{1}{c}{$J_3$} & 
			\multicolumn{1}{c}{$J_4$} \\
			\noalign{\vskip 3pt}
			\colrule
			$\text{EuCd}_2\text{P}_2$                    & 0.3035 & -0.3518 & \multicolumn{1}{c}{---} & \multicolumn{1}{c}{---} \\
			$\text{Eu}_{0.67}\text{Gd}_{0.33}\text{Cd}_2\text{P}_2$ & 3.453  &  5.166  & 1.44                    & 0.964  \\
			$\text{Eu}_{0.5}\text{Gd}_{0.5}\text{Cd}_2\text{P}_2$  & 0.663  & -0.0131 & 1.461                   & \multicolumn{1}{c}{---} \\
			$\text{Eu}_{0.33}\text{Gd}_{0.67}\text{Cd}_2\text{P}_2$ & 2.70  &  1.066  & 6.856                    & 4.170 \\
			$\text{GdCd}_2\text{P}_2$                    & 1.526  &  1.699  & \multicolumn{1}{c}{---} & \multicolumn{1}{c}{---} \\
			                  
		\end{tabular}
	\end{ruledtabular}
\end{table}

\begin{figure*}[t]
	\centering
	\includegraphics[width=17cm]{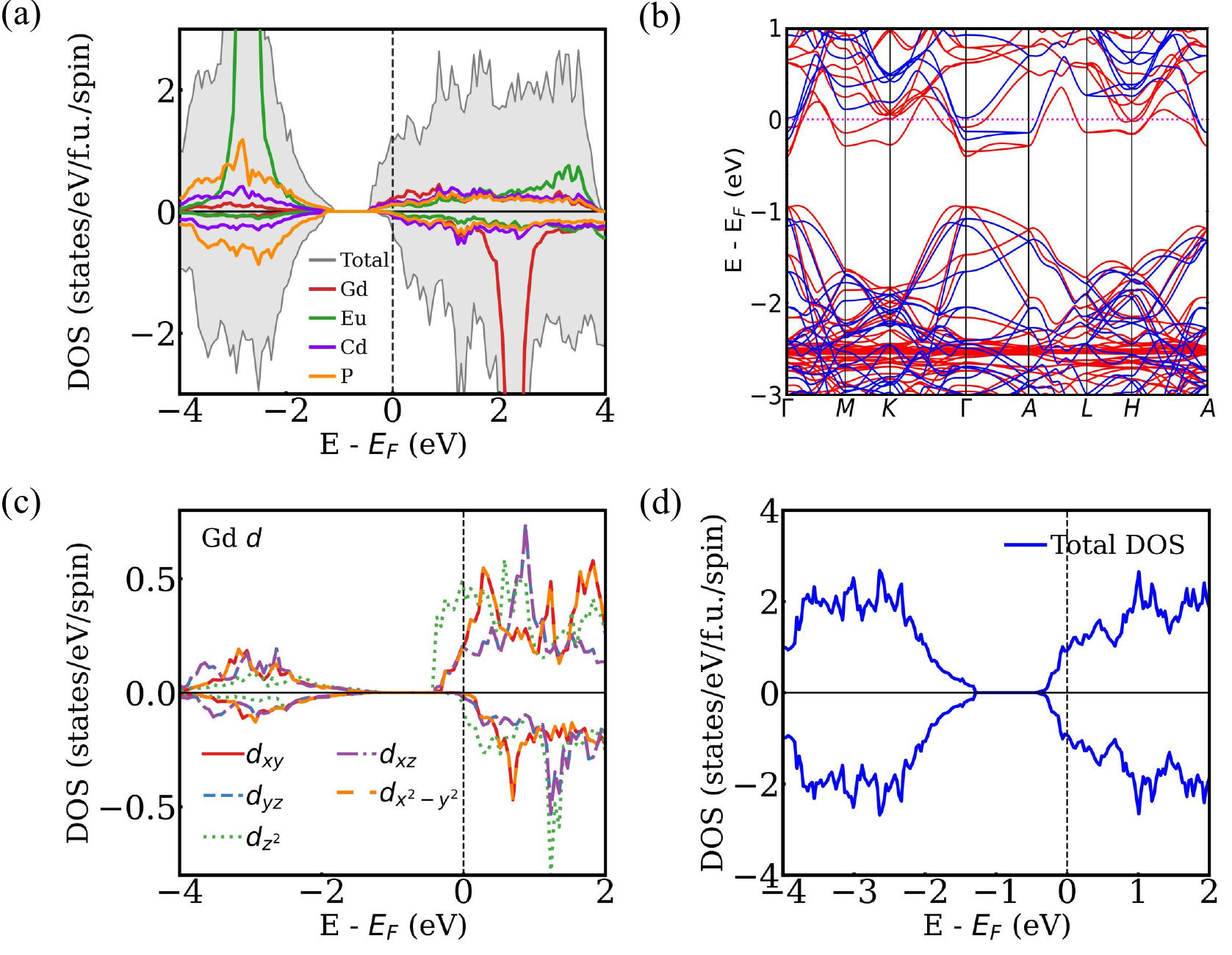}
	\caption{Electronic structure of $\text{Eu}_{0.67}\text{Gd}_{0.33}\text{Cd}_2\text{P}_2$ in the FM state: (a) PDOS, (b) band structure, and (c) lm-decomposed $d$-orbital DOS for Gd, all calculated within $\text{GGA} + U$; and (d) total density of states (TDOS) in the paramagnetic (PM) state calculated within standard GGA, treating 4$f$ electrons as core states.}
	\label{Fig5}
\end{figure*}
\begin{figure*}[t]
	\centering
	\includegraphics[width=17cm]{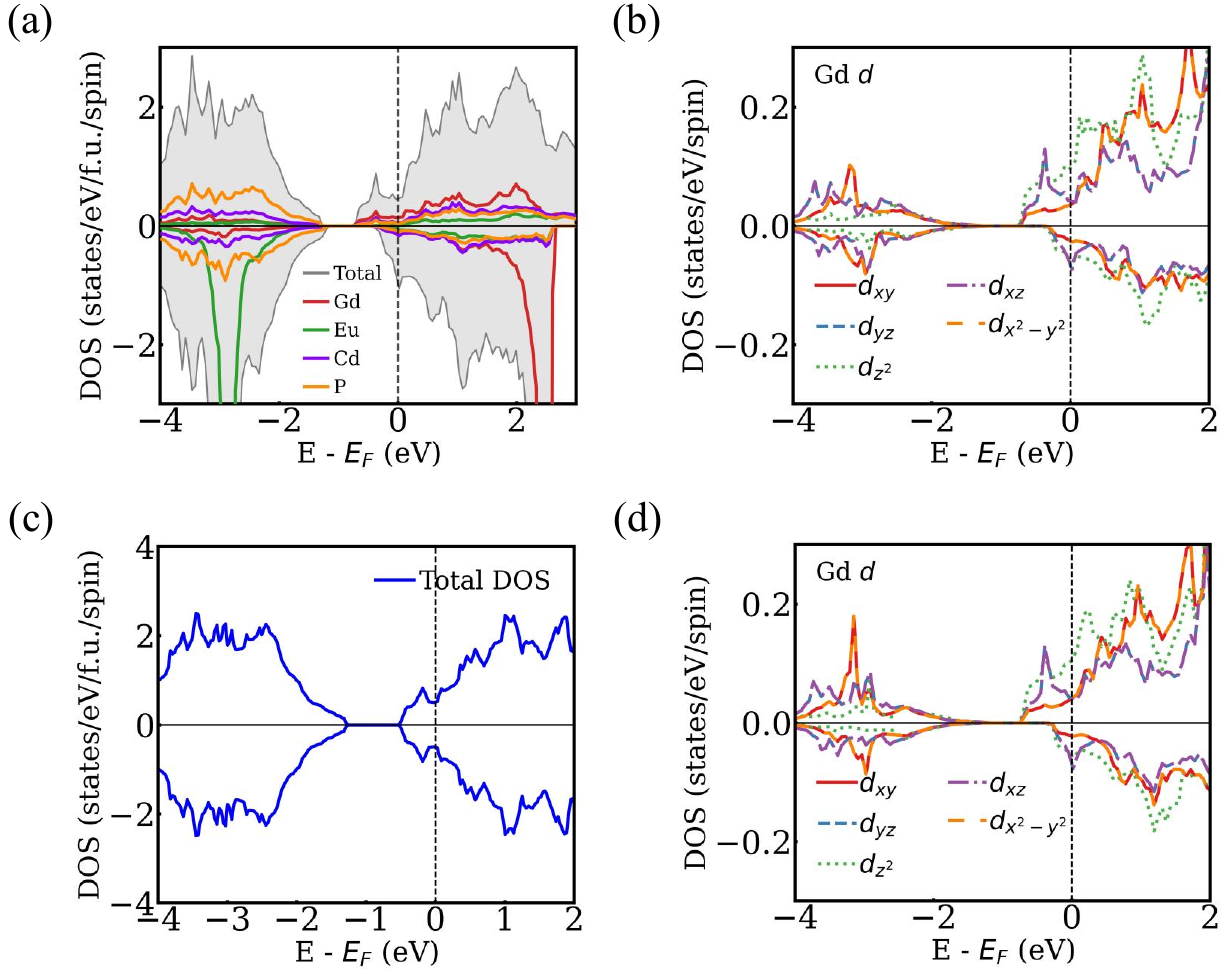}
	\caption{Electronic density of states of Eu$_{0.5}$Gd$_{0.5}$Cd$_2$P$_2$: (a) PDOS in the A-AFM state within $\text{GGA} + U$, (b) lm-decomposed $d$-orbital DOS for Gd in the A-AFM state within $\text{GGA} + U$, (c) TDOS in the PM state calculated within standard GGA, treating 4$f$ electrons as core states, and (d) lm-decomposed $d$-orbital DOS for Gd in the FM state within $\text{GGA} + U$.}
	\label{Fig6}
\end{figure*}

\begin{figure*}[t]
	\centering
	\includegraphics[width=17cm]{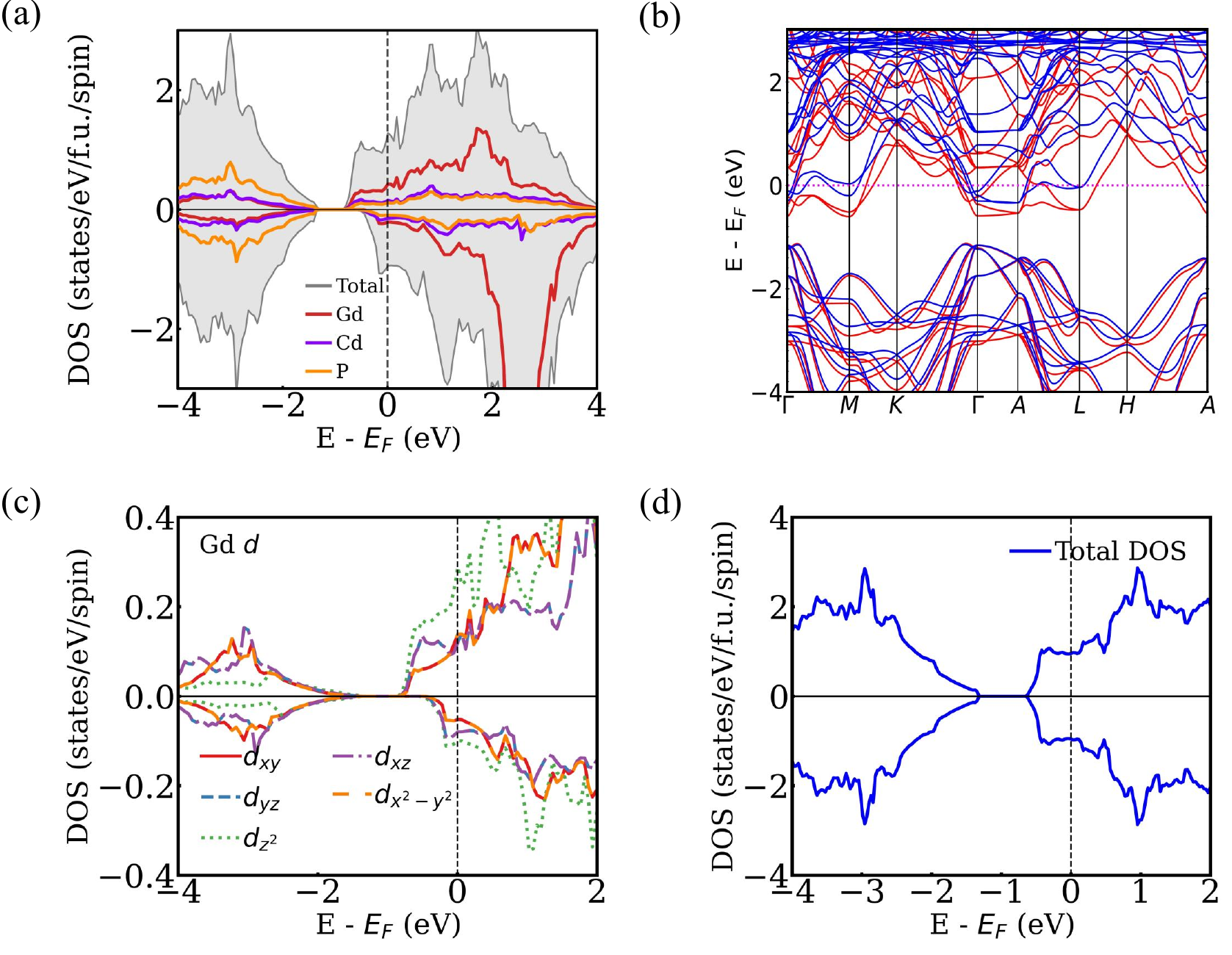}
	\caption{Electronic structure of $\text{Gd}\text{Cd}_2\text{P}_2$ in the FM state: (a) PDOS, (b) band structure, and (c) lm-decomposed $d$-orbital DOS for Gd, all calculated within $\text{GGA} + U$; and (d) TDOS in the PM state calculated within standard GGA, treating 4$f$ electrons as core states.}
	\label{Fig7}
\end{figure*}
\begin{figure}[t]
	\centering
	\includegraphics[width=8.5cm]{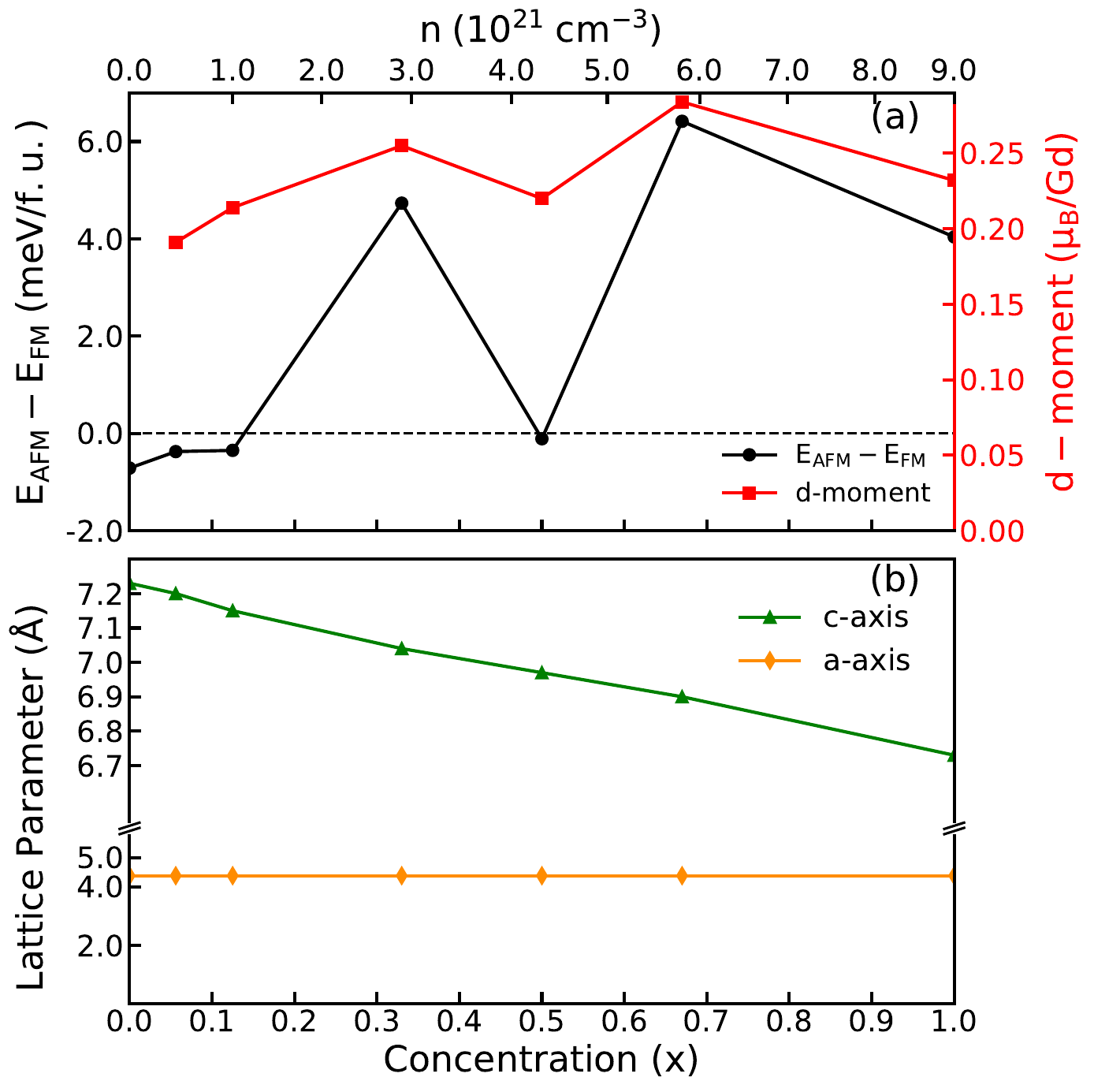}
	\caption{Fig. represent (a) the energy difference ($\Delta E$) and $d$-moment per Gd ion, and (b) the lattice parameters $a$ and $c$, as a function of Gd concentration.}
	\label{Fig8}
\end{figure}
\begin{figure}[t]
	\centering
	\includegraphics[width=8.5cm]{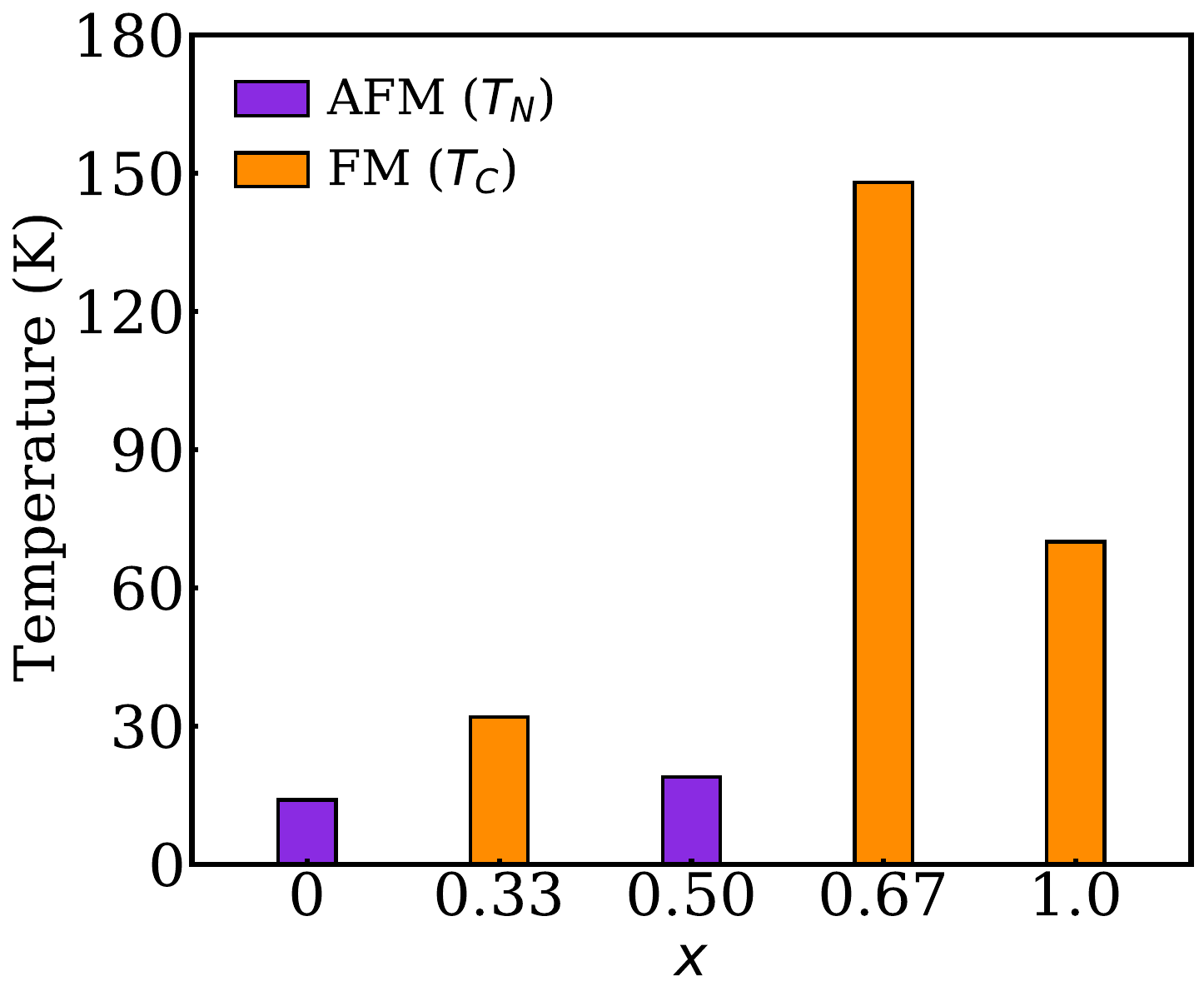}
	\caption{Histogram representing Gd concentration (x) vs transition temperature (T).}
	\label{Fig9}
\end{figure}
\begin{figure*}[htbp!]
	\centering
	\includegraphics[width=17cm]{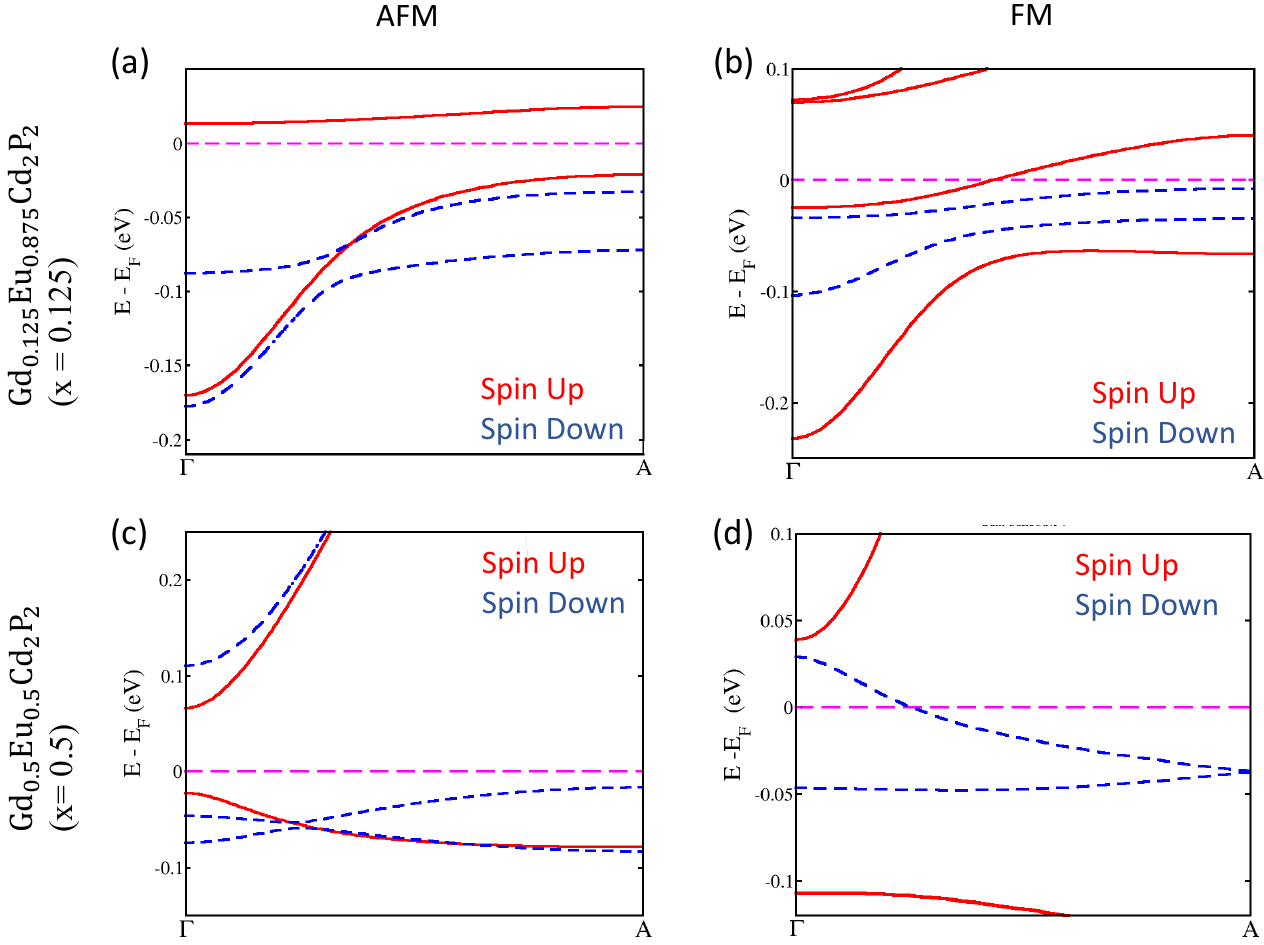}
	\caption{Comparision of the band crossing along the $\Gamma$--$A$ $k$-path of the FM and AFM states for x=0.125 and x=0.5 cases, calculated within $\text{GGA} + U$ approximation.}
	\label{Fig10}
\end{figure*}


\subsection{$\mathbf{Eu_{0.67}Gd_{0.33}Cd_2P_2}$}
\vspace{-2mm}
To achieve 33 $\%$ concentraion of Gd, we consider a $\sqrt{3} \times \sqrt{3} \times 1$ supercell of EuCd$_2$P$_2$ (see Fig.~\hyperref[Fig1]{\ref{Fig1}(b)}) and replaced 1/3rd of Eu ions by Gd ions. The total energy calculations presented in Table~\ref{TableI} reveal that the magnetic ground state for x = 0.33 is ferromagnetic (FM) whereas the parent compound and the doped compounds for x = 0.056 and 0.125 were A-AFM. We further observe that the Eu and Gd spin moments prefer to stay in the $ab$-plane. The electronic structure calculations (see DOS and band structure calculated with GGA+U in Fig~\ref{Fig5} and GGA+U+SO in Fig. S2 in Supplemental Materials) show that the compound becomes metallic in nature. This is because the Fermi level is now straddled by the hybridized states constituted by Gd 5\textit{d}, Eu 5\textit{d}, Cd \textit{s} and P \textit{p} states (see band characters plotted in Fig. S3 in Supplemental Materials). The extra electron coming from Gd 5\textit{d} predominantly occupies the $\textit{d}_{z^2}$ orbital (see Fig.~\hyperref[Fig5]{\ref{Fig5}(c)} on orbital projected \textit{d} DOS) at the Fermi level. There is a strong spin splitting also seen between up and down spin channels of \textit{d}-states at the Fermi level giving rise to a strong spin polarization. This spin polarization is induced by the strong ferromagnetic 4\textit{f}-5\textit{d} exchange as we do not see any spin polarization when we consider 4\textit{f} states frozen in core (see Fig.~\hyperref[Fig5]{\ref{Fig5}(d)}). The magnetic moment due to \textit{d}-electron is found to be 0.255~$\mu_B$.\par  
We have calculated the exchange constants ($J_1$, $J_2$, $J_3$ and $J_4$ as depicted in Fig.~\hyperref[Fig1]{\ref{Fig1}(b)}) which are listed in Table~\ref{TableII}. Using these exchange constants we have performed Monte-Carlo simulations on Heisenberg model to calculate the temperature dependence of specific heat, which we have presented in Fig.~\hyperref[Fig4]{\ref{Fig4}(b)}. As one can see, the transition temperature for the FM phase is around 32 K.

\subsection{$\mathbf{Eu_{0.5}Gd_{0.5}Cd_2P_2}$}
\vspace{-2mm}
We replace a single Eu in $1\times1\times2$ supercell of EuCd$_2$P$_2$ with one Gd to get 50 $\%$ doping concentration of Gd. Before evaluating the magnetic ground state, we have calculated the total energies of different site ordering of Gd and Eu ions. We observe that the lowest energy structure has Gd and Eu ions segregate in successive $ab$-planes along the $c$-direction (see Fig.~\hyperref[Fig1]{\ref{Fig1}(c)}). Phonon calculations further confirmed the dynamical stability of this structure (see Fig.~\hyperref[Fig2]{\ref{Fig2}(c)}). Regarding the magnetic structure, our calculations reveal that A-AFM is favored over FM state in Eu$_{0.5}$Gd$_{0.5}$Cd$_2$P$_2$ even though the energy difference between A-AFM and FM states is very small (see Table~\ref{TableI}). This indicates the presence of competing A-AFM and FM phases in this compound. Our GGA+$U$+SO calculations with $U_{\text{eff}}$ = 7 eV further reveal that A-AFM with magnetic moments along the $a$-direction is the magnetic ground state (see Table~\ref{TableI}). 

Electronic structure calculations reveal a metallic nature for both A-AFM and FM states with Fermi level dominantly occupied by Gd \textit{d} states as shown in Figs.~\hyperref[Fig6]{\ref{Fig6}} and in Fig. S4 in Supplemental materials. However, unlike the previous case of x = 0.33, the Gd \textit{d} electron occupies dominantly $\textit{d}_{z^2}$, $\textit{d}_{xz}$ and $\textit{d}_{yz}$ orbitals (see Fig.~\hyperref[Fig6]{\ref{Fig6}(b)}). This could possibly explain why FM interaction along the $c$-direction is not favored in this case, whereas dominant occupancy of $\textit{d}_{z^2}$ orbital might be responsible for the FM interaction along the $c$-direction in x = 0.33 case. The band characters are plotted in Fig. S5 and S6 for A-AFM and FM respectively. 

Three exchange constant ($J$) values \textit{viz.}, $J_1$, $J_2$ and $J_3$ were estimated (see Table~\ref{TableII}), among which $J_1$ and $J_3$ are intralayer interaction between Eu-Eu and Gd-Gd, respectively, whereas $J_2$ denotes the interlayer interaction between Gd-Eu (see Fig.~\hyperref[Fig1]{\ref{Fig1}(c)}). The A-AFM transition temperature ($T_\text{N}$) of Eu$_{0.5}$Gd$_{0.5}$Cd$_2$P$_2$ calculated from Monte-Carlo simulations is found to be around 19 K, as seen Fig.~\hyperref[Fig4]{\ref{Fig4}(c)}, where the specific heat ($C_\text{V}$) versus Temperature (T) plot is presented.

\subsection{$\mathbf{Eu_{0.33}Gd_{0.67}Cd_2P_2}$}
\vspace{-2mm}
Considering the same supercell of EuCd$_2$P$_2$ (see Fig.~\hyperref[Fig1]{\ref{Fig1}(b)}) that was used for x = 0.33, we have created the compound with x = 0.67 by replacing 2/3rd of the Eu ions with Gd ions. The total energy calculations presented in Table~\ref{TableI} reveal that the magnetic ground state for x = 0.67 is ferromagnetic (FM), which is the most stable with respect to the A-AFM state among all the compounds studied here. In this case, the Eu and Gd spin moments also prefer to stay in the $ab$-plane. The electronic structure (see DOS and band structure in Fig. S8 in the supplementary material) is very similar to that observed for x = 0.33. 

Calculated the exchange constants ($J_1$, $J_2$, $J_3$ and $J_4$) are listed in Table~\ref{TableII}. Our Monte-Carlo simulations on Heisenberg model reveal a high transition temperature $\approx{150}$ K (highest among all the compounds studied here) for the ferromagnetic state as can be seen in the temperature dependence of specific heat presented in Fig.~\hyperref[Fig4]{\ref{Fig4}(d)}.

\subsection{{$\mathbf{GdCd_2P_2}$}}
\vspace{-2mm}
Replacing all the Eu ions by Gd ions (see Fig.~\hyperref[Fig1]{\ref{Fig1}(c)}), we created a new compound GdCd$_2$P$_2$ which is not experimentally synthesized yet. Our phonon calculations on fully relaxed structure clearly show that this compound is dynamically stable (see Fig.~\hyperref[Fig2]{\ref{Fig2}(d)}) and hence it should be feasible to prepare it experimentally in future. Calculations within GGA+$U$ with $U_{\text{eff}}$ = 7 eV show that FM state has lower energy than A-AFM in this case. 

Further, GGA+$U$+SO calculations show that FM (100) (i.e. FM state with Gd moments pointing along the $a$-direction) is found to be magnetic ground state. The electronic structure presented in Fig.~\ref{Fig7} shows a strong spin splitting of Gd d-states induced by a strong 4f-5d exchange in the FM state which is not present in PM state. We also see that the Fermi level is dominantly occupied by d$_{z^2}$ orbital. Calculated exchange constants (see Table~\ref{TableII}) clearly shows ferromagnetic nature. Transition temperature ($T_\text{C}$) of GdCd$_2$P$_2$ was found to be around 70 K, as indicated by Fig.~\hyperref[Fig4]{\ref{Fig4}(e)} of the specific heat ($C_\text{V}$) versus Temperature (T) plot simulated using Monte-Carlo method. 


%


\section{\textbf{Discussion}} 
\label{SecIV}
\vspace{-2mm}

In Sec.~\ref{SecIII}, we have comprehensively discussed all the calculations and results which include structural stability, ground state magnetic order, electronic band structures, density of states (DOS) and magnetic transition temperatures from specific heat calculation. This section is dedicated to drawing key inferences from the above mentioned results. Gd substitution at the Eu site in EuCd$_2$P$_2$ at various concentrations results in a wide range of carrier densities~(n), ranging from $10^{20}\text{ to }10^{22}\, \mathrm{ electrons\,cm^{-3}}$ (see Fig.~\ref{Fig8}). One of the important observations presented above is the oscillation of ground state magnetic order between A-AFM and FM as a function of Gd concentration (see Fig.~\hyperref[Fig8]{\ref{Fig8}(a)} and the phase diagram presented in Fig.~\ref{Fig9} as a histogram plot of magnetic transition temperature versus Gd concentration). At lower Gd concentrations (such as x = 0.056, 0.125) A-AFM remains the ground state same as the parent compound EuCd$_2$P$_2$ though the energy difference between A-AFM and FM ($\Delta E = E_{\text{AFM}} - E_{\text{FM}}$) decreases by half from that observed in parent compound (see Table~\ref{TableI}). However, as we increase Gd concentration further, at x = 0.33 the ground state becomes FM with large energy difference ($\Delta E$) which gives rise to higher exchange constants (Table~\ref{TableII}) and consequently a higher FM transition temperature ($T_\text{C}$) (see Fig.~\ref{Fig9}). Surprisingly, at x = 0.5 the ground state becomes A-AFM with very small $\Delta E$ suggesting a competing FM excited state. However, on further increasing Gd concentration (such as x = 0.67 and x = 1), the ground state stabilizes as FM with high magnetic transition temperatures (see Fig.~\ref{Fig9}). In Fig.~\hyperref[Fig8]{\ref{Fig8}(b)}, we present the variation of lattice constants with Gd concentration (x). We observe that though $a$ and $b$ do not depend significantly on x, $c$ decreases almost linearly with x. Thus, the interlayer separation along $c$ decreases with increasing Gd substitution which is expected to increase the FM $T_\text{C}$ as we observe here. This has been seen experimentally in EuZn$_2$P$_2$ \cite{hole_doped_EuZn2P2}. In all the Gd substituted compounds conduction states are found to be spin polarized due to strong onsite ferromagnetic 4f-5d exchange interaction as indicated by a finite d-moment (see Fig.~\hyperref[Fig8]{\ref{Fig8}(a)}).

To understand the oscillating ground state magnetic order behavior, let us look into the electronic structure of the compound as a function of Gd concentration. The ground state of the parent compound EuCd$_2$P$_2$ is observed to be A-AFM semiconductor with a gap of about 724 meV and a transition temperature of 14 K (see Figs.~\hyperref[Fig3]{\ref{Fig3}(b)}, and~\hyperref[Fig4]{\ref{Fig4}(a)}). With the substitution of Gd the ground state becomes A-AFM metal up to x = 0.125 with Gd $d$ states primarily at the Fermi level. With further doping we find that the compound transits to a FM metallic ground state at x = 0.33 with $T_\text{C}$ around 32 K. As can be seen in PDOS, the Fermi level is predominantly occupied by the $d_{z^2}$ orbital of Gd 5$d$ which drives the FM exchange along $c$ direction. A strong 4\textit{f}-5\textit{d} exchange is observed in both Gd and Eu states causing a high spin polarization of particularly Gd \textit{d} states. This is clearly seen when we compare the Gd \textit{d} states for paramagnetic phase (which was simulated by freezing the 4\textit{f} states in the core). In paramagnetic state there is no spin polarization of Gd/Eu \textit{d} states (see Figs.~\hyperref[Fig5]{\ref{Fig5}(d)},~\hyperref[Fig6]{\ref{Fig6}(c)}, and~\hyperref[Fig7]{\ref{Fig7}(d)}) whereas in FM/A-AFM phase a strong spin polarization is observed (see Figs.~\hyperref[Fig5]{\ref{Fig5}(a)},~\hyperref[Fig6]{\ref{Fig6}(a)}, and~\hyperref[Fig7]{\ref{Fig7}(a)}). This also has implications for magnetotransport properties which we will discuss below. Interestingly, as we doped further at x = 0.5 we observe A-AFM metallic ground state again with $T_\text{N}$ = 19 K. Looking at the PDOS of Gd \textit{d} states we observe that three d-orbitals (i.e. $d_{z^2}$, $d_{xz}$ and $d_{yz}$) are predominantly occupied. This may possibly explain why the exchange interaction along $c$ direction becomes AFM in this case. As we proceed further with Gd substitution, we observe FM metallic state for x = 0.33, x = 0.67 and x = 1 with itinerant electron in the $d_{z^2}$ orbital dominating the Fermi level and thereby driving the FM exchange interaction along $c$. We observe highest $T_\text{C}$ of 150 K for x = 0.67 and also highest spin polarization reflected in Gd \textit{d} magnetic moment as plotted in Fig.~\hyperref[Fig8]{\ref{Fig8}(a)}. 

The spin polarization caused by the 4\textit{f}-5\textit{d} exchange in FM and A-AFM states is expected to give rise to magnetoresistance around the magnetic transition when one comes from the high temperature paramagnetic state where up and down spin channels are equally occupied. Random scattering in the paramagnetic state causing high resistivity will be much reduced due to strong spin polarization of the conduction electrons in FM or AFM phase. With the application of magnetic field one is expected to observe further increase in the spin polarization leading to enhanced magnetoresistance in this compound. However, this mechanism is similar to what is expected in a typical FM metal. Below we discuss two cases where a dramatic change in resistance is expected to occur along the $c$ direction on the application of external magnetic field (i.e., colossal anisotropic magnetoresistance).  

At x = 0.125 and x = 0.5, the ground state magnetic order is A-AFM, but the FM state lies very close in energy. On comparing the band structures along $\Gamma$--$A$ direction (see Fig.~\ref{Fig10}) we observe that in both cases, A-AFM has a band gap at the Fermi level whereas in FM, a band crosses the Fermi level (i.e., FM state becomes conducting along the $c$ direction, while the A-AFM was insulating). A clear difference is also observed in the Fermi surfaces of A-AFM and FM states presented in Fig. S7 in the supplemental material. Given the low energy difference between these two states, one can expect to turn the A-AFM insulating state along the $c$ direction to FM metallic state using an external magnetic field easily. This will lead to colossal anisotropic magnetoresistance in this compound for x = 0.125 and x = 0.5 along the $c$ direction. Recently, Acharya \textit{et al}.~\cite{GMR_GdPS} underscored the crucial role of strong on-site $f\text{--}d$ exchange coupling in Gd atoms as the origin of gigantic isotropic magnetoresistance in GdPS. On the other hand, Wang \textit{et al}.~\cite{AMR_EuZn2As2} probed the microscopic origin of magnetic and transport anisotropy through a comparative study of electronic band structures in related layered Zintl pnictides such as $\text{EuZn}_2\text{As}_2$. 

The underlying mechanism can be compared with that operating in electron doped manganites~\cite{TM_2002, TM_2003_PRB}, where the double exchange mechanism without the dominant Jahn-Teller effect is considered. It has been shown that in A, C, and G-type AFM phases, it is the orbital order that drives the transport properties~\cite{TM_2003_PRB}. In our compound, though, we don't have a mixed valency state like manganites; Gd$^{2+}$/Eu$^{2+}$ create a similar situation to that of Mn$^{3+}$/Mn$^{4+}$, with localized 4$f^7$ spins at Gd/Eu sites and an itinerant Gd \textit{d} electron occupying the hybridized Gd/Eu 5\textit{d} states, which are coupled by a strong 4\textit{f}-5\textit{d} ferromagnetic exchange. 


In summary, we have performed a systematic investigation of the effect of electron doping via Gd substitution on the magnetic transition, electronic structure and magnetoresistive properties of Eu$_{1-\text{x}}$Gd$_{\text{x}}$Cd$_{2}$P$_{2}$ for x = 0, 0.056, 0.125, 0.33, 0.5, 0.67 and 1.0. Though Gd substition led to a metallic state for all x values studied, the ground state magnetic order is found to oscillate between A-AFM and FM states. While stable FM metallic states are observed with high transition temperatures for x=0.33, 0.67 and 1, A-AFM metallic ground state with competing FM excited state is observed for x=0.056, 0.125 and 0.5. Highest FM transition temperature of about 150 K is observed for x=0.67. Most importantly, the A-AFM ground state observed for x=0.5 and 0.125 are insulating along $c$ direction while the FM excited state (which is very close in energy) is metallic along the same direction. Therefore, we predict that a dramatic change in resistance will be observed along $c$ direction on the application of external magnetic field (i.e. colossal anisotropic magnetoresistance) for x=0.5 and 0.125. Underlying mechanism can be similar to the double exchange mechanism proposed for electron doped manganites which is driven by the orbital order rather than the Jahn-Teller effect. Eu$^{2+}$/Gd$^{2+}$ mimic the mixed valence of manganites. With a strong onsite 4f-5d ferromagnetic exchange and an itinerant electron coming from Gd 5d in the hybrizied conduction band brings in a double exchange like scenario. Future experiments can shed more light on these compounds. 

\begin{acknowledgments}
	\vspace{-2.5mm}
	YSN acknowledges the Ministry of Education, India, for research fellowship. TM and YSN acknowledge the National Supercomputing Mission (NSM) for providing computing resources of `PARAM Ganga' at the Indian Institute of Technology Roorkee, which is implemented by C-DAC and supported by the Ministry of Electronics and Information Technology (MeitY) and the Department of Science and Technology (DST), Government of India.
\end{acknowledgments}


\end{document}